\documentclass[pra,twocolumn,floatfix,showpacs,tightenlines]{revtex4}
\usepackage{graphicx}

\begin{document}

\title{Effect of anisotropic hopping on the Bose Hubbard 
model phase diagram: strong-coupling perturbation theory on a square lattice} 
\author{J.~K.~Freericks}
\affiliation{Department of Physics, Georgetown University, Washington, DC
20057, U.S.A.}

\date{\today}

\begin{abstract}
There has been a recent resurgence of experimental efforts to quantitatively determine the phase diagram of the Bose Hubbard model by carefully analyzing experiments with ultracold bosonic atoms on an optical lattice.  In many realizations of these experiments, the hopping amplitudes are not homogeneous throughout the lattice, but instead, the lattice has an anisotropy where hopping along one direction is not exactly equal to hopping along a perpendicular direction. In this contribution, we examine how an anisotropy in the hopping matrix elements affects the Mott lobes of the Bose Hubbard model.  For weak anisotropy, we find the phase diagram is only slightly modified when expressed in terms of the average hopping, while for strong anisotropy, one expects to ultimately see dimensional crossover effects.
\end{abstract}


\maketitle

\section{Introduction}. 

Recent efforts in ultracold atomic gases have been looking at the prospects for building an optical lattice emulator---namely an analog quantum simulator that can model simplified many-body Hamiltonians used in condensed matter physics, and extract important physical properties of the models, like their phase diagrams~\cite{feynman,jaksch}. In particular, the recent experiments of Spielman~\cite{spielman}, {\it et al.}, determined the critical interaction strength for the loss of superfluidity in the Bose Hubbard model on a square lattice. Their measured value is within 6\% of the theoretically calculated values for the critical point ($U_c\approx 15.8(20)$ in the experiment at a finite temperature versus $U_c\approx 16.74$ in theoretical calculations at $T=0$).

The optical lattice formed in this experiment was constructed by a single laser beam that was reflected to form standing waves in two different directions, creating a square lattice. Due to geometrical constraints in the experimental set up, the hopping matrix elements differed in the $x$ and $y$ directions (due to slightly different potential well depths in the two directions).  The question we want to answer here is how much does the location of the critical point depend on the hopping anisotropy?  Obviously, if the anisotropy is so large that the system becomes weakly coupled one-dimensional chains, then we expect to see a dimensional crossover of the critical behavior at the tip of the lobe; in this case it would change from the power law behavior in two dimensions to a Kosterlitz-Thouless behavior in one dimension. Having a solid quantitative understanding of the importance of the anisotropy is critical for analyzing any given experiment.

Since the critical $U$ values are large for the Mott insulator to superfluid transition, a strong-coupling analysis is the most appropriate way to proceed.  Strong coupling expansions have been shown to be very accurate in determining the phase diagram~\cite{strong1,strong2,elstner_monien}, especially when they are coupled with a scaling analysis of the tip of the Mott phase lobe. It is a straightforward exercise to modify the strong-coupling formulas to take into account the anisotropy of the hopping matrix elements and determine their effect on the phase diagram. In this contribution, we will examine the effect of anisotropy on the two-dimensional square lattice.  This is the first nontrivial system where anisotropy plays a role, and hence it should be the simplest to analyze as an explicit example. The other reason to look at this case is that experiments have already been performed, so we can quantitatively determine what the effects of anisotropy are on those experiments. 

From a heuristic standpoint, it is obvious that we should express results in terms of the average hopping matrix element $\bar t=(t_x+t_y)/2$, because only the average matrix element enters into the first-order shift of the energies of the particle or hole defect phases of the Mott insulator (see below).  Hence the deviations due to an anisotropy, which will depend on $\Delta t=(t_x-t_y)$ enter first quadratically in $(\Delta t)^2$.  This result follows from the fact that we should not be able to tell the difference of a system with a slightly larger hopping in the $x$ direction versus one with a slightly smaller hopping in the $x$ direction, because the two cases can be mapped onto each other by rotating the lattice by 90 degrees (if we always use the hopping in the $y$ direction as our reference hopping). So an anisotropy of 15\% should produce corrections on the order of 2\%.  Such corrections are likely to be smaller than other experimental errors, so the effect of anisotropy is expected to be minor for realistic experimental situations.
Indeed, this is the main conclusion of this work, but we actually find that the tip of the Mott phase lobe is even less sensitive to the effects of anisotropy so the correction factor is much smaller.

In Section II, we derive the expressions for the strong-coupling analysis of the Mott insulator and of the particle and hole defect phases and show how the phase boundaries are modified by the anisotropy.  In Section III, we present numerical results and a scaling analysis of the behavior of the critical point and how it changes with anisotropy. Conclusions follow in Section IV.

\section{Formalism}. 

We begin with the Bose Hubbard model whose Hamiltonian is~\cite{fishers}
\begin{equation}
  \mathcal{H} = - \sum_{ij} t_{ij} b^\dagger_i b^{\phantom{\dagger}}_j
   - \mu \sum_i {n_i} + \frac{1}{2} U \sum_i {n}_i ({n}_i-1),
  \label{eq: H}
\end{equation}
where $b_i$ is the boson
annihilation operator at site $i$, $n_i=b_i^\dagger b_i^{}$ is the bosonic number operator at site $i$, $-t_{ij}$ is the hopping matrix element
between the site $i$ and $j$, 
$U$ is the strength of the on-site repulsion, and $\mu$ is the chemical
potential. The hopping matrix is nonzero for nearest neighbors on a two-dimensional square lattice. The hopping is allowed to be anisotropic, so that $t_x=t_{ii+x}$ corresponding to hopping parallel to the $x$-axis need not be equal to $t_y=t_{ii+y}$, corresponding to hopping parallel to the $y$-axis.  The lattice does maintain translation invariance though, so all hopping matrix elements along the $x$-direction are equal to $t_x$ and similarly for the $y$-direction. It is more convenient to express the results in terms of the average hopping matrix element 
\begin{equation}
 \bar t=\frac{t_x+t_y}{2},
\label{eq: t_ave}
\end{equation}
and the relative hopping matrix element
\begin{equation}
 \Delta t=t_x-t_y.
\label{eq: t_rel}
\end{equation}
Using these definitions it is easy to see that
\begin{equation}
 t_x^2+t_y^2=2\bar t^2 +\frac{1}{2} (\Delta t)^2,
\label{eq: t_quad}
\end{equation}
and
\begin{equation}
 t_x^3+t_y^3=2\bar t^3+\frac{3}{2}\bar t (\Delta t)^2.
\label{eq: t_cub}
\end{equation}
These relations will become important below.

We will be examining the stability of the Mott insulating phase with $n_0$ bosons per lattice site. Following Ref.~\cite{strong2}, we need to evaluate the perturbative energy for adding one particle or for adding one hole into the Mott phase.  To zeroth order in the hopping, the ground state is $N$-fold degenerate, with $N$ the number of lattice sites on the lattice.  The hopping matrix breaks this degeneracy (to first order in the hopping).  We need to find the lowest eigenvalue and eigenvector of the single particle matrix $S_{ij}=-t_{ij}$ to perform the degenerate perturbation theory.  Since the lattice is translationally invariant, the lowest eigenvector is the vector $(1,1,1,...,1)/\sqrt{N}$ with an eigenvalue $\lambda_{\rm min}=-2t_x-2t_y=-4\bar t$. The chemical potential within the Mott phase is written as the deviation from integral multiples of $U$: $\mu=(n_0+\delta)U$. The point $\mu=n_0 U$ corresponds to the upper tip of the $n_0$ Mott phase lobe, which meets the lower tip of the $n_0+1$ Mott phase lobe. The variable $\delta$ is negative for a particle defect and positive for a hole defect.

Under the assumption that the compressibility vanishes continuously at the Mott phase lobe boundary, the transition between the incompressible (Mott) phase and the compressible
(superfluid) phase is determined when the energy difference between the Mott insulator
and the defect state (particle or hole) vanishes.  
The two branches of the Mott-phase boundary meet when
\begin{equation}
  \delta^{\rm (part)}(n_0) + 1 = \delta^{\rm (hole)}(n_0).
  \label{eq: critical condition}
\end{equation}
The additional one on the left hand side
arises because $\delta$ is measured from the point
$\mu/U = n_0$, and the hole defect for the $n_0$ Mott insulator starts at $\mu/U=n_0-1$.
Equation (\ref{eq: critical condition})
is used to estimate the critical value of
the hopping matrix element beyond which the Mott-insulator phase ceases to exist.  

The energy of the Mott phase satisfies~\cite{strong1,strong2}
\begin{eqnarray}
 \frac{E_{\rm Mott}(n_0)}{N}&=&-\delta U n_0-\frac{U}{2}n_0(n_0+1)\nonumber\\
&-&\frac{4\bar t^2+(\Delta t)^2}{U}n_0(n_0+1),
\label{eq: energy_mott}
\end{eqnarray}
through third order in the hopping matrix element. Notice that we used the definition in Eq.~(\ref{eq: t_quad}) and that no odd powers in the hopping enter because the square lattice is bipartite. 

Similarly, the energy to add a particle to the system satisfies~\cite{strong1,strong2}
\begin{widetext}
 \begin{eqnarray}
  E_{\rm{Def}}^{\rm (part)}(n_0) &-& E_{\rm{Mott}}(n_0) =
  -\delta^{(\rm{part})} U-4\bar t(n_0+1)
  +\frac{4\bar t^2+(\Delta t)^2}{2U}n_0(5n_0+4)
  -\frac{16\bar t^2}{U}n_0(n_0+1)\nonumber\\
  &+&\frac{1}{U^2} n_0(n_0+1)
  \Biggr[-64(2n_0+1)\bar t^3+(25n_0+14)
  \bar t[4\bar t^2+(\Delta t)^2]-2(2n_0+1)[4\bar t^3+3\bar t(\Delta t)^2]\Biggr]
  \label{eq: Edef upper}
 \end{eqnarray}
while the energy to add a hole is
\begin{eqnarray}
 E_{\text{Def}}^{\rm (hole)}(n_0) &-& E_{\rm{Mott}}(n_0)=
  \delta^{(\rm{hole})}U-4\bar t n_0
  + \frac{4\bar t^2+(\Delta t)^2}{2U}(n_0+1)(5n_0+1)
  -\frac{16\bar t^2}{U}n_0(n_0+1)\nonumber\\
  &+&\frac{1}{U^2} n_0(n_0+1)
  \Biggr[-64(2n_0+1)\bar t^3+(25n_0+11)\bar t   [4\bar t^2+(\Delta t)^2] -2(2n_0+1)[4\bar t^3+3\bar t(\Delta t)^2]\Biggr].
  \label{eq: Edef lower}
\end{eqnarray}
\end{widetext}
Both results hold through third-order perturbation theory in the hopping.
We take each equation, find the point where the defect energy is equal to the Mott energy (i.e., the LHS vanishes), and then solve for the critical chemical potential $\delta$ as a function of $\bar t$ and $\Delta t$.

Note that we can obtain an estimate of the value of $\Delta t/\bar t$ needed for the dimensional crossover because the Mott phase lobes in one dimension have an opposite sign curvature from that in two dimensions, due to the cusp-like behavior of the Kosterlitz-Thouless transition.  So a simple estimate for the dimensional crossover is when the coefficient of the quadratic term in the hopping vanishes for the energy differences of the defect phases.  When $n_0=1$, this occurs at $t_y/t_x=8-\sqrt{63}\approx 0.0627$ for the particle branch of the lobe and $t_y/t_x=2-\sqrt{3}\approx 0.2679$ for the hole branch of the lobe, while the change in curvature occurs at $t_y/t_x=4-\sqrt{15}\approx 0.1270$ for both branches when we take the limit $n_0\rightarrow\infty$. This implies that we can crudely estimate the dimensional crossover to occur when $t_y/t_x$ is on the order of $0.1$--$0.05$, or $\Delta t/\bar t\approx 1.6$--$1.8$.

We will use scaling theory to extrapolate the short-term power series expansion into a functional form appropriate for the Mott phase lobe. To begin, we define the parameter $x$ by
\begin{equation}
 x=\frac{2\bar t}{U},
\label{eq: xdef}
\end{equation}
and then express the boundaries of the Mott phase lobe branches via a power series that includes the known power-law critical behavior of the tip of the lobe~\cite{fishers}
\begin{eqnarray}
 \delta^{\rm part}&=&A+Bx+Cx^2+Dx^3+...\nonumber\\
&+&(x_c-x)^{Z\nu}(\alpha+\beta x+
\gamma x^2+...),\label{eq: delta_x_part}
\\
-1+\delta^{\rm hole}&=&A+Bx+Cx^2+Dx^3+...\nonumber\\
&-&(x_c-x)^{Z\nu}(\alpha+\beta x+
\gamma x^2+...),\label{eq: delta_x_hole}
\end{eqnarray}
where the constants $A$, $B$, $C$, $D$, ... and $\alpha$, $\beta$, $\gamma$, ... can depend on $\Delta t/\bar t$ and $Z\nu$ is the critical exponent which determines the shape of the Mott phase lobe near the critical point $x_c$.
We immediately find
\begin{eqnarray}
 A&=&-\frac{1}{2},\label{eq: A}\\
B&=&-1,\label{eq: B}\\
C&=&-\frac{1}{4}(2n_0+1)\left [ 1+\frac{1}{4}\left ( \frac{\Delta t}{\bar t}\right )^2\right ],\label{eq: C}\\
D&=&\frac{3}{4}n_0(n_0+1)\left [ 1+\frac{1}{4}\left ( \frac{\Delta t}{\bar t}\right )^2\right ].\label{eq: D}
\end{eqnarray}
The expression for the critical behavior is somewhat more complicated:
\begin{eqnarray}
 &~&(x_c-x)^{Z\nu}(\alpha+\beta x+\gamma x^2)=\frac{1}{2}-x(2n_0+1)\nonumber\\
&~&~-\frac{x^2}{4}\left [ 6n_0^2+6n_0-1-\left ( 5n_0^2+5n_0+\frac{1}{2}\right ) \left ( \frac{\Delta t}{\bar t}\right )^2\right ]\nonumber\\
&~&~+\frac{x^3}{4}n_0(n_0+1)(2n_0+1)\left [ -11+\frac{13}{4}\left (\frac{\Delta t}{\bar t}\right )^2\right ].
\label{eq: critical}
\end{eqnarray}
Note that the equality in Eq.~(\ref{eq: critical}) is meant only in the sense that it holds term by term in $x^n$ when the LHS is expanded in a power series in $x$ for small $x$.

Performing the Taylor series expansion of the power law for small $x$ and equating terms (keeping the series up
to only the $\gamma$ term) produces the following four equations that the coefficients, $x_c$ and $Z\nu$ satisfy:
\begin{widetext}
\begin{eqnarray}
 x_c^{Z\nu}\alpha&=&\frac{1}{2};\label{eq: alpha}\\
\frac{1}{2}\left ( \frac{\beta}{\alpha}-\frac{Z\nu}{x_c}\right )&=&-(2n_0+1);\label{eq: beta}\\
\frac{1}{2}\left ( \frac{\gamma}{\alpha}-\frac{Z\nu\beta}{\alpha x_c}+\frac{Z\nu(Z\nu-1)}{2x_c^2}\right )
&=&-\frac{1}{4}\left [ 6n_0^2+6n_0-1-\left ( 5n_0^2+5n_0+\frac{1}{2}\right ) \left (\frac{\Delta t}{\bar t}\right )^2\right ];
\label{eq: gamma}\\
\frac{1}{2}\left ( -\frac{Z\nu\gamma}{\alpha x_c}+\frac{Z\nu(Z\nu-1)\beta}{2\alpha x_c^2}
-\frac{Z\nu(Z\nu-1)(Z\nu-2)}{6x_c^3}\right )&=&\frac{1}{4}n_0(n_0+1)(2n_0+1)\left [ -11+\frac{13}{4}\left (\frac{\Delta t}{\bar t}\right )^2\right ].\label{eq: delta}
\end{eqnarray}
\end{widetext}
There are two ways to perform the scaling analysis.  In an unconstrained analysis, we set $\gamma=0$ and solve Eqs.~(\ref{eq: alpha}--\ref{eq: delta}) for $Z\nu$, $x_c$, $\alpha$, and $\beta$.  In a constrained analysis, we assume the three-dimensional XY model critical behavior holds for all $\Delta t/\bar t$ values of interest, and so we set $Z\nu=0.67$ and fit $x_c$, $\alpha$, $\beta$, and $\gamma$.  Both of these techniques will be compared numerically
in the next section.

\section{Results}. 

\begin{figure}[htb]
\centerline{\includegraphics [width=3.3in, angle=0]  {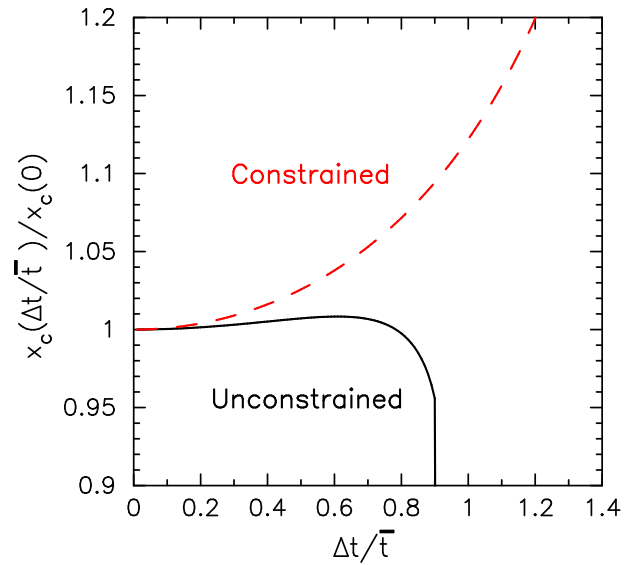}}
\caption[]{(Color online.) Plot of the relative change of the location of the tip of the $n_0=1$ Mott phase lobe as a function of the anisotropy.  The solid (black) line is for the unconstrained scaling approach and the dashed (red) line is for the constrained scaling approach.  Note how the two curves agree for small anisotropy and how the effect is quite small.  When the anisotropy becomes large enough, the series is to small to carry out the scaling analysis and the results become inaccurate.
}
\label{fig: xc}
\end{figure}

We focus on the $n_0=1$ Mott phase lobe since this has been the most accurately measured phase diagram of the model.  For the unconstrained scaling analysis, we set $\gamma=0$ and solve for $Z\nu$, $x_c$, $\alpha$, and $\beta$. Using $\alpha=1/2x_c^{Z\nu}$, $\beta=[-6+(Z\nu/x_c)]\alpha$, and defining
$y=Z\nu/x_c$ produces the following two equations to self-consistently solve for the critical point and exponent
\begin{eqnarray}
 y&=&6-\frac{1}{2x_c}\label{eq: unconstrained1}\\
&+&\frac{1}{2}\sqrt{(12-x_c)^2+44-42\left (\frac{\Delta t}{\bar t}\right )^2},\nonumber\\
\frac{1}{x_c}&=&\frac{9}{2}\label{eq: unconstrained2}\\
&+&\frac{1}{y}\sqrt{y^4-9y^3+\frac{81}{4}y^2+99y-\frac{117}{4}y\left (\frac{\Delta t}{\bar t}\right )^2}.\nonumber
\end{eqnarray}
These equations are most simply solved by iteration.  Start with $x_c=0.1$ and iteratively solve Eq.~(\ref{eq: unconstrained1}) for $y$ and then Eq.~(\ref{eq: unconstrained2}) for $x_c$ until they stop changing.   A plot of the relative change of the critical point $x_c(\Delta t/\bar t)/x_c(0)$ versus the anisotropy $\Delta t/\bar t$ is shown in Fig.~\ref{fig: xc} with the solid line. We find the critical point initially increases until $\Delta t\approx 0.6 \bar t$ where it begins to decrease. But, the effect is much smaller than one would have naively guessed.  For example, if $\Delta t/\bar t=0.1$, we find $x_c(\Delta t/\bar t)/x_c(0)=1.0004$, which is $0.04$ times smaller than the crude estimate that the deviation is quadratic in the anisotropy. 

These equations can no longer be solved once $\Delta t > 1.02\bar t$, where the critical $x_c$ is pushed to zero and the exponent vanishes.  Obviously the scaling approach cannot faithfully determine the Mott phase lobe when $\Delta t/\bar t$ becomes too large.  In actual experiments, we expect the anisotropy to be small (on the order of 15\% or less) so this is not a big issue. Surprisingly, the critical exponent $Z\nu$ {\it decreases} with anisotropy.  One might have expected it to increase to approach the cusp-like behavior of the one-dimensional system, but it doesn't do that.  Perhaps this is because one needs higher order terms in the series to properly estimate the exponent~\cite{elstner_monien}. Note that at $\Delta t/\bar t=0$, we have $x_c=0.1121$ and $Z\nu=0.5826$, while the true critical point and exponent satisfy~\cite{elstner_monien,svistunov} $x_c=0.1195$ and $Z\nu=0.67$. While this truncated analysis is accurate to about 7\% for the critical point in the absence of anisotropy, we expect the truncated analysis to be much more accurate for the relative change of the critical point with anisotropy, especially for small anisotropy.  Indeed, the increase of the critical $x$ for small anisotropy is precisely what we would expect, since the Mott phase lobe in one dimension has a cusplike shape and is pushed to larger $x$ values than in two dimensions, so we expect the anisotropy to move the critical point to the right in the phase diagram.

Next, we examine the constrained scaling theory analysis, where we fix $Z\nu=0.67$ and we solve for $x_c$, $\alpha$, $\beta$, and $\gamma$. In this case, we have a cubic equation to solve for $x_c$ which is simple to write down by eliminating $\alpha$, $\beta$ and $\gamma$ from the coupled equations for the critical point. The solution of these equations is also plotted in Fig.~\ref{fig: xc} with the dashed line. Note how the critical value of $x_c$ always increases when we constrain the exponent; the method fails for $\Delta t/\bar t>1.75$.  The constrained analysis is much more accurate for the critical point, predicting it to lie at $x_c=0.1173$ which is only a 2\% error.  The change of $x_c$ with anisotropy continues to be small, but not as small as in the unconstrained analysis; we find $x_c(\Delta t/\bar t)/x_c(0)=1.0010$ at  $\Delta t/\bar t=0.1$, which is still a factor of ten smaller than we originally estimated. But, the trend is once again to increase the critical $x$ value, which is just what we expect when we compare the one-dimensional and two-dimensional Mott phase lobes.

It appears that the constrained extrapolation procedure estimates the change in the anisotropy better than the unconstrained approach.  However, in both cases, the change in the critical point due to anisotropy is much too small to be observed in experiment for anisotropies on the order of 15\% or less.  But, it is imperative that one uses the average hopping as the energy scale in order for the corrections to be small.  If any other relative unit for hopping is chosen, the changes in the critical point can be very large (linear in the anisotropy).

\section{Conclusions and Discussion}

We have examined the effect of hopping anisotropy on the shape of the Mott phase lobe of a two-dimensional square lattice.  When results are expressed in terms of the average hopping, we find the critical point changes by an amount that is too small to be seen experimentally for reasonable experimental parameters (an anisotropy of 15\% say). This implies that one need not worry about controlling the lattice parameters to be identical when examining the Mott phase transition in the Bose Hubbard model. 

One can ask what will happen in three dimensions?  Once again, if we express results in terms of the average hopping, we expect the errors to scale quadratically in the anisotropy.  Furthermore, since the shape of the Mott phase lobe for two dimensions and for three dimensions is quite similar, we anticipate that the effect of anisotropy would be even smaller for the three dimensional case. Hence, we do not believe the effect of anisotropy will lead to any significant modification of experimental results on these systems as long as the anisotropy is kept at the modest level of about 15\% or less.

\acknowledgments  
We acknowledge useful conversations with T.-L. Ho, W. D. Phillips, J. V. Porto, and I. B. Spielman.
This work was supported under ARO Award W911NF0710576 with funds from the DARPA OLE Program.


\begin{thebibliography}{99}
\bibitem{feynman}
R. P. Feynman, Int. J. Theor. Phys. {\bf 21}, 467 (1982).
\bibitem{jaksch}
D. Jaksch, C. Bruder, J. I. Cirac, C. W. Gardiner, and P. Zoller, Phys. Rev. Lett. {\bf 81}, 3108 (1998)
\bibitem{spielman}
I. B. Spielman, W. D. Phillips, and J. V. Porto, Phys. Rev. Lett. {\bf 98}, 080404 (2007); Phys. Rev. Lett. {\bf 100}, 120402 (2008).
\bibitem{strong1}
J. K. Freericks and H. Monien, Europhys. Lett.  {\bf 26}, 545-550 (1994).
\bibitem{strong2}
J. K. Freericks and H. Monien, Phys. Rev. B  {\bf 53}, 2691 (1996).
\bibitem{elstner_monien}
N. Elstner and H. Monien, Phys. Rev. B {\bf 59}, 12184 (1999); and eprint arXiv:cond-mat/9905367.
\bibitem{fishers}
M. P. A. Fisher, P. B. Weichman, G. Grinstein, and D. S. Fisher, Phys. Rev. B {\bf 40}, 546 (1989).
\bibitem{svistunov}
B. Capogrosso-Sansone, S. G. S\"oyler, N. Prokof'ev, and B. Svistunov, Phys. Rev. A {\bf 77}, 015602 (2008).

\end{thebibliography}
\end{document}